\documentclass[runningheads]{llncs}
\usepackage[T1]{fontenc}
\usepackage{graphicx}

\usepackage{caption}
\usepackage{subcaption}
\usepackage{booktabs}
\usepackage{multirow}
\usepackage{array}
\newcolumntype{L}[1]{>{\raggedright\let\newline\\\arraybackslash\hspace{0pt}}m{#1}}
\newcolumntype{C}[1]{>{\centering\let\newline\\\arraybackslash\hspace{0pt}}m{#1}}
\newcolumntype{R}[1]{>{\raggedleft\let\newline\\\arraybackslash\hspace{0pt}}m{#1}}

\usepackage[breaklinks,colorlinks]{hyperref}

\usepackage{color}

\begin{document}
\title{Metrics to Quantify Global Consistency in Synthetic Medical Images}
\titlerunning{Metrics to Quantify Global Consistency}
\author{Daniel Scholz\inst{1,2,*}
\and
Benedikt Wiestler \inst{2}
\and
Daniel Rueckert\inst{1,3}
\and
Martin J. Menten\inst{1,3}
}
\authorrunning{D. Scholz et al.}
 \institute{
 Lab for AI in Medicine, Technical University of Munich, Germany  \\
 \and Department of Neuroradiology, Klinikum rechts der Isar, Technical University of Munich, Germany \\
 \and BioMedIA, Department of Computing, Imperial College London, UK \\
  $^*$\email{daniel.scholz@mri.tum.de}
}
\maketitle

\begin{abstract}
Image synthesis is increasingly being adopted in medical image processing, for example for data augmentation or inter-modality image translation.
In these critical applications, the generated images must fulfill a high standard of biological correctness.
A particular requirement for these images is global consistency, i.e an image being overall coherent and structured so that all parts of the image fit together in a realistic and meaningful way. 
Yet, established image quality metrics do not explicitly quantify this property of synthetic images.
In this work, we introduce two metrics that can measure the global consistency of synthetic images on a per-image basis.
To measure the global consistency, we presume that a realistic image exhibits consistent properties, e.g., a person's body fat in a whole-body MRI, throughout the depicted object or scene.
Hence, we quantify global consistency by predicting and comparing explicit attributes of images on patches using supervised trained neural networks.
Next, we adapt this strategy to an unlabeled setting by measuring the similarity of implicit image features predicted by a self-supervised trained network.
Our results demonstrate that predicting explicit attributes of synthetic images on patches can distinguish globally consistent from inconsistent images.
Implicit representations of images are less sensitive to assess global consistency but are still serviceable when labeled data is unavailable.
Compared to established metrics, such as the FID, our method can explicitly measure global consistency on a per-image basis, enabling a dedicated analysis of the biological plausibility of single synthetic images.

\keywords{Generative Modeling \and Synthetic Images \and Image Quality Metrics \and Global Consistency}
\end{abstract}

\section{Introduction}

\begin{figure}[htpb]
     \centering
         \includegraphics[width=\linewidth]{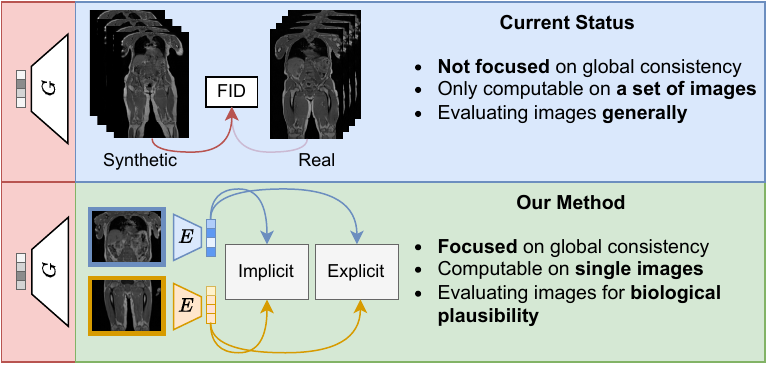}
         \caption{
         We present a novel method to quantify global consistency in generated images.
        Most established image quality metrics, like FID~\cite{heusel2017gans}, are not designed to measure the biological correctness of medical images.
        Conversely, our approach measures the global consistency of synthetic medical images, like whole-body MRIs, based on their explicit and implicit features.}
         \label{fig:whole-body-mri-example}
\end{figure}

With recent improvements in deep learning-based generative modeling~\cite{ho2020denoising,sauer2022stylegan,dhariwal2021diffusion}, image synthesis is increasingly utilized in medical image processing.
It has seen application in inter-modality transfer~\cite{liu2021ct}, counterfactual image generation~\cite{menten2023exploring}, anomaly detection~\cite{han2021madgan}, data augmentation~\cite{sun2022hierarchical}, and synthetic dataset generation~\cite{pinaya2022brain}.
When using synthetic images in critical medical systems, it is vital to ensure the biological correctness of the images.
One crucial aspect of image realism is its global consistency \cite{esser2021taming,johnson2018image,zhang2019self}.
Global consistency refers to an image's overall coherence and structure so that all parts of the image fit together in a realistic and plausible way.
While several others have researched methods to improve the global consistency of synthetic images \cite{hudson2021ganformer,liang2022nuwainfinity}, these works do not quantitatively assess the global consistency of these images in a standardized fashion.
This is because existing metrics, such as Inception Score \cite{salimans2016improved}, Fréchet Inception Distance (FID) \cite{heusel2017gans}, and Precision and Recall \cite{sajjadi2018assessing,kynkaanniemi2019improved}, only measure image quality in terms of fidelity and variety.

In this work, we introduce solutions to measure the global consistency of synthetic images.
To this end, we make the following contributions.

\begin{itemize}
    \item We propose an approach to quantify global consistency by determining attributes on different image regions.
    We call this method \emph{explicit} quantification of global consistency.
    \item Next, we adapt this approach to a setting in which explicit labels are not available.
    To this end, we utilize the cosine similarity between feature vectors of patches in the image as a global consistency measure.
    These \emph{implicit} features are predicted by neural networks trained in a self-supervised fashion.
    \item In extensive experiments, we compare our proposed metrics with FID, one of the most established image quality metrics, with regard to its ability to measure global consistency in synthetic images.
    We perform our experiments on the challenging task of whole-body magnetic resonance image (MRI) synthesis, in which it is crucial that the various depicted body parts match.
\end{itemize}

\section{Related Works}

\subsection{Global Consistency in Image Synthesis}
The notion of global consistency in image synthesis has been researched in computer vision.
Multiple important works \cite{johnson2018image,zhang2019self} describe synthesizing complex images with multiple objects as challenging and lacking global coherence.
Integrating the attention mechanism \cite{NIPS2017_3f5ee243} into the GAN architecture \cite{hudson2021ganformer,esser2021taming} facilitates generating more globally consistent images.
To evaluate their adherence to the properties in the real data, Hudson \mbox{\textit{et al.}}~\cite{hudson2021ganformer} statistically compare property co-occurrences in the generated images, similar to \cite{sun2022hierarchical}.
The use of large auto-regressive models advances the generation of ultra-high-resolution images while maintaining global consistency \cite{liang2022nuwainfinity}.
They use a block-wise FID to assess the quality of individual blocks in the image, which only evaluates the realism of individual patches but does not measure the global consistency within a single image. 
In summary, none of these works have dedicated quantitative metrics for global consistency.

\subsection{Metrics Measuring Quality of Generated Images}
Several metrics, such as Inception Score~\cite{salimans2016improved}, Fréchet Inception Distance~(FID)~\cite{heusel2017gans}, and Precision and Recall~\cite{kynkaanniemi2019improved,sajjadi2018assessing}, have been proposed in the literature to assess the quality of synthetic images.
The most established metric, the FID \cite{heusel2017gans}, measures image quality and variation in a single value by comparing the distribution over features from sets of real and synthetic images. 
Multiple variants have been proposed in the literature to address the limitations of FID. 
These variants focus on overcoming the bias towards a large number of samples \cite{bińkowski2018demystifying,chong2020effectively}, the lack of spatial features \cite{Tsitsulin2020The} or standardization of its calculation \cite{parmar2021cleanfid}.
However, the global consistency remains, at most, only validated as part of general image fidelity.

Zhang \textit{et al.}~\cite{zhang2018unreasonable} measure a learned perceptual image patch similarity (LPIPS) between patches of two separate images.
While this metric is conceptually similar to ours, their work focuses on evaluating different kinds of representations and similarity measures between two images for their correspondence to human judgment.
However, they do not assess global consistency within a single image.
Sun \textit{et al.} \cite{sun2022hierarchical} evaluate their hierarchical amortized GAN by quantifying the accuracy of clinical predictions on synthetic images.
Their evaluation strategy only compares statistics over the clinical predictions between real and synthetic data but does not incorporate per-image analysis.
In general, existing metrics do not explicitly address the quantification of global consistency.

\subsection{GANs for Whole-body MRI Synthesis}
Only few works have researched the challenging task of generating synthetic whole-body MRIs. 
Mensing \textit{et al.} \cite{mensing20223d} adapt a FastGAN \cite{liu2021towards} and a StyleGAN2 \cite{hong20213d,karras2020analyzing} to generate whole-body MRIs.
They primarily evaluate their generated images using the Fréchet Inception Distance (FID) \cite{heusel2017gans}.
However, they do not focus on assessing global consistency of the synthetic images.

\section{Method}
\begin{figure}
    \centering
    \includegraphics[width=\linewidth]{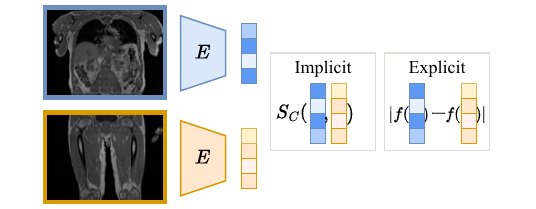}
    \caption{
    Two strategies to assess the global consistency of an image based on the feature representations of the superior and inferior half of the body.
    \emph{Explicit}: Absolute error between an explicit attribute predicted from the feature representation using some regression head $f$.
    \emph{Implicit}: Cosine similarity $S_C$ between the feature representations.
    }
    \label{fig:method-metric}
\end{figure}

We propose two novel metrics to measure the global consistency of synthetic images.
We distinguish between \emph{implicit} and \emph{explicit} quantification of global consistency, which are described in the following (see Figure \ref{fig:method-metric}).

\subsection{Explicit Quantification} \label{sec:explicit-quantification}
Our method for explicitly quantifying global consistency is based on the notion that biological properties should be consistent in different parts of a person's body.
For example, a person's body mass index (BMI) should be similar when viewing the superior part of a whole-body MRI depicting the torso and the inferior part containing the legs.
To assess its global consistency, we compare various biological attributes, such as age, body fat percentage, or BMI, in two parts of the synthetic images.
While individual organs might age at different rates \cite{schaum2020ageing}, our method assumes that the overall age of the superior part and inferior part of a person's body still contain consistent age-related information.
In addition, the body fat mass between the limbs and the trunk correlates and can hence serve as marker for consistency in a synthetic image~\cite{jung2021relationship}.
We generate two views of the whole-body MRI by simply cropping the superior and inferior half of the image.
Other possible cropping modes include random cropping, cropping based on semantic regions, such as organs, and grid-structured cropping.
The two views are each evaluated using dedicated \emph{referee neural networks}.
We train several neural networks in a supervised fashion to predict one of three different biological properties for either the superior or inferior image view.

By comparing the predicted attributes via the absolute error, we can obtain a proxy measure for the global consistency of a synthetic image.
For a more holistic analysis, we simultaneously compare an average error of all biological attributes.

\subsection{Implicit Quantification}
Detailed annotations of the data are not always available, rendering supervised training of referee networks infeasible.
Therefore, we propose the use of implicit features extracted via a network that has been trained via self-supervision as an alternative measure for global consistency.

As before, we crop two views from the synthetic image and extract one feature vector for each view by applying an encoder network.
Here, the encoder network is trained using SimCLR~\cite{chen2020simple}, a self-supervised contrastive learning framework alleviating the need for labels during training.
SimCLR is trained to return similar representations for two views of the same image and diverging representations for two views of different images.
The similarity between the embedding of the two views is obtained by calculating their cosine similarity.
To calculate a global consistency measure for a given image, we obtain the cosine similarity between the embeddings of the superior and inferior views.

\subsection{Experimental Setup}
We conduct experiments using 44205 whole-body MRIs from the UK Biobank population study \cite{sudlow2015uk}, which we split into 36013 training images, 4096 validation images, and 4096 test images.
We extract the slice along the coronal plane in the intensity center of mass of the 3d volumes and normalize them to the range of $[0,1]$.
We train one ResNet50~\cite{he2016deep} network per attribute on the training set as a referee network for the explicit quantification experiments.
We also fit a ResNet50 using SimCLR~\cite{chen2020simple} to our training images to extract features for the implicit quantification strategy.

The validation images are used to evaluate the accuracy of the referee networks for the explicit quantification strategy.
We find that the networks achieve good performance on the attribute estimation.
The mean absolute error (MAE) for age estimation is 3.9 years $\pm$ 2.98 years on the superior half and 4.4 years $\pm$ 3.35 years on the inferior half.
Similarly, we achieve an MAE of 0.97 $\pm$ 0.83 on the superior and 1.11 $\pm$ 0.93 on the inferior half for BMI estimation and 2.10\% $\pm$ 1.70\% on the superior and  2.36\% $\pm$ 1.89\%  on the inferior half for body fat percentage prediction.
Ultimately, we compare the variation in biological properties of the explicit metric, the cosine similarity of the implicit metric, and the FID on all test set images.

\section{Results}

\subsection{Distinguishing Consistent From Inconsistent Images}
\begin{figure}[!ht]
    \centering
    \begin{subfigure}{0.3\textwidth}
        \centering
         \includegraphics[height=0.9\linewidth, angle=0]{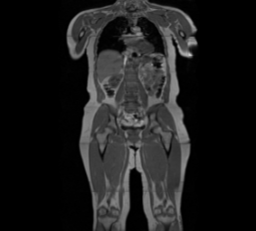}
    \end{subfigure}%
    \hfill
    \begin{subfigure}{0.3\textwidth}
    \centering
         \includegraphics[width=0.9\linewidth, angle=90]{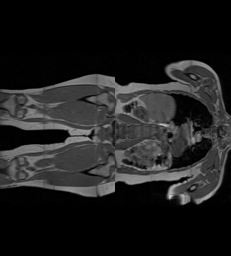}
    \end{subfigure}%
    \hfill
    \begin{subfigure}{0.3\textwidth}
        \centering
         \includegraphics[width=0.9\linewidth, angle=90]{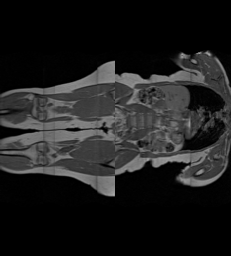}
    \end{subfigure}%
    \caption{A comparison of an original whole-body MRI (left) with the modified versions used in our experiments, i.e, consistent (middle) and inconsistent (right) superior-inferior combinations with the central 10\% removed.}
    \label{fig:consistent-inconsistent-images}
\end{figure}

Initially, we analyze the two proposed metrics on a dataset of consistent and inconsistent images.
We construct the inconsistent images by stitching the superior part and inferior part of two different whole-body MRIs from the test set together (see Figure~\ref{fig:consistent-inconsistent-images}).
The sharp edge at the seam of the inconsistent images is a very distinctive feature.
In order to avoid the metrics being influenced by it, we remove the central 10\% of both the consistent and inconsistent images.

We compare our two metrics with the FID~\cite{heusel2017gans}, which is calculated using two distinct sets of images.
One half of the consistent images serves as the reference dataset for calculating the FID of either the other half of the consistent images or the inconsistent images, respectively.

Our metrics differentiate well between consistent and inconsistent images (see Table~\ref{tab:experiments}, top).
For the explicit strategy, we report the mean over the superior-inferior errors of age, BMI, and body fat percentage prediction after normalizing them to a range between 0 and 1.
While the FID is also influenced by global consistency, our metric distinguishes more clearly between consistent and inconsistent.

\begin{table}
    \centering
    \caption{
Comparison of our explicit global consistency metrics, implicit global consistency metric, and FID in two different experiments.
In the first one, we calculate all metrics on constructed consistent and inconsistent images.
In the second experiment, the metrics are compared for real and synthetic datasets, akin to the envisioned use case of our proposed method.
} \label{tab:experiments}
\begin{tabular}{@{}l C{1.2cm}C{3cm}C{2.5cm}@{}}
\toprule
Dataset & FID ($\downarrow$) & Explicit (Ours, $\downarrow$) & Implicit (Ours, $\uparrow$) \\ \midrule
Consistent       & \textbf{14.10}                         & \textbf{0.09 $\pm$ 0.05}            & \textbf{0.59 $\pm$ 0.12}                       \\
Inconsistent    & 16.10                                  & 0.24 $\pm$ 0.11                     & 0.37  $\pm$ 0.17                               \\ \midrule
Real                                 & \textbf{14.10}                           &  0.09 $\pm$ 0.050           & \textbf{0.59 $\pm$ 0.12}                       \\
Synthetic  & 17.13                                    & 0.09 $\pm$ 0.049                    & 0.55 $\pm$ 0.14                                \\ \bottomrule
    \end{tabular}
\end{table}

\begin{figure}[!ht]
    \centering
   \includegraphics[width=0.9\linewidth]{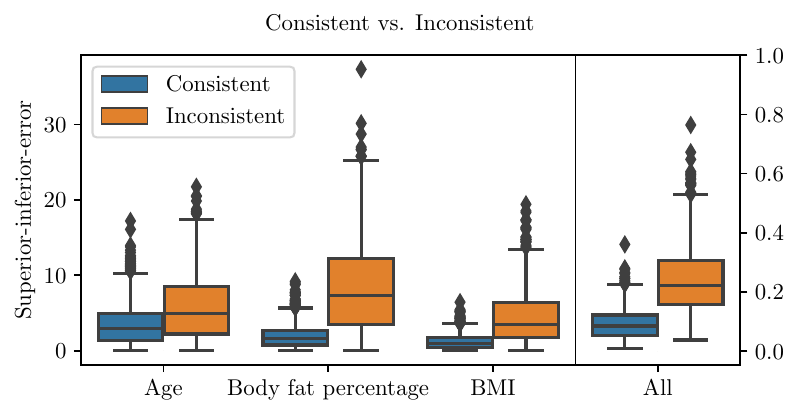}
   \includegraphics[width=0.9\linewidth]{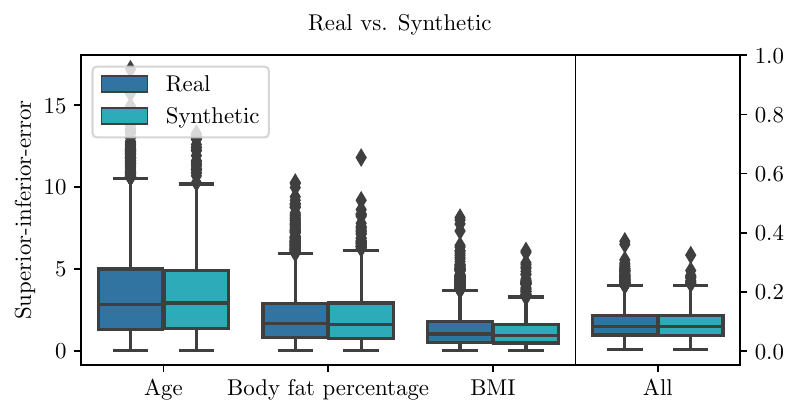}
    \caption{
    The per-attribute results of the explicit absolute errors between the superior and inferior part of the consistent and inconsistent images (top) and real and synthetic images (bottom).
    The rightmost column: an average over the 0-1-normalized per-attribute errors.
    } \label{fig:explicit-boxplot_real}
\end{figure}

We present a detailed analysis of the explicit attribute errors in Figure~\ref{fig:explicit-boxplot_real}.
The experiment shows that body fat percentage and BMI are more distinctive biological attributes than age.

\begin{figure}[!ht]
    \centering
    \includegraphics[width=\textwidth]{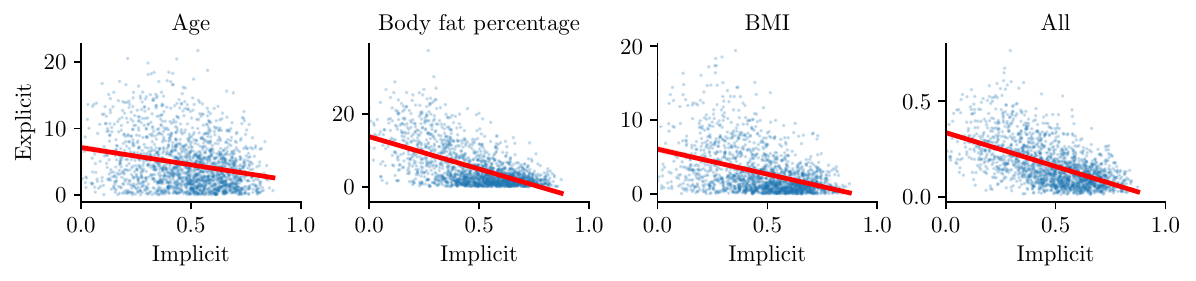}
    \caption{
    The correlations between our implicit and explicit metrics verifying the utility of the implicit strategy in the absence of labels on real images.
    } \label{fig:explicit-implicit-correlation-constructed}
\end{figure}

Additionally, we investigate the correlation between our implicit and explicit metrics to verify the utility of the implicit strategy in the absence of labels (see Figure~\ref{fig:explicit-implicit-correlation-constructed}).
These findings suggest the potential utility of the implicit quantification strategy as a weaker alternative to explicit quantification.

\subsection{Global Consistency in Synthetic Whole-Body MRIs}

We conduct an exemplary assessment of global consistency on 1000 synthetic images using our implicit and explicit metrics and the FID.
The synthetic whole-body MRIs were generated using a StyleGAN2~\cite{karras2020analyzing} that we trained on images of the UK Biobank~\cite{sudlow2015uk}.
The results suggest an overall high global consistency and a low error in biological attributes in the synthetic images (Table~\ref{tab:experiments}, bottom).
The images show overall high fidelity to the real images due to the comparable FID to the real images.

Our metrics differ only slightly between real and synthetic in the per-attribute analysis (see Figure~\ref{fig:explicit-boxplot_real}, bottom).
The high values in our metrics indicate a high degree of global consistency in the synthetic images.

\section{Discussion and Conclusion}
In this work, we have proposed two strategies to quantify the global consistency in synthetic medical images.
We found that global consistency influences established metrics for synthetic image quality, such as the FID, yet the differences between consistent and inconsistent images are more pronounced in our novel metrics.
Our first metric explicitly quantifies the error between predicted biological attributes in the superior and inferior half of a single whole-body MR image.
However, this approach relies on labels to train neural networks that determine the biological attributes.
As a solution, we also presented a second metric based on implicit representations that can be obtained via a self-supervised trained network.
Both strategies have proven suitable for assessing synthetic medical images in terms of their biological plausibility.

We envision that our work will complement the existing landscape of image quality metrics - especially in medical imaging - and that it will be used to develop and benchmark generative models that synthesize globally consistent medical images.
An extension of our work to the 3D domain is theoretically simple but may be practically challenging due to the additional complexity when training \mbox{SimCLR} for the implicit and the referee networks for the explicit metric.
Moreover, global consistency analysis for other image modalities and use cases can be enabled through retraining the feature extraction networks on domain specific data with corresponding augmentations.
Ultimately, we believe our work can potentially increase the trust in using synthetic data for critical medical applications in the future.

\subsubsection{Acknowledgments}
This research has been conducted using the UK Biobank Resource under Application Number 87802.
This work is funded by the Munich Center for Machine Learning.

\bibliographystyle{splncs04}
\bibliography{references}

\end{document}